\documentclass[fleqn,usenatbib]{mnras}

\usepackage[T1]{fontenc}

\DeclareRobustCommand{\VAN}[3]{#2}
\let\VANthebibliography\thebibliography
\def\thebibliography{\DeclareRobustCommand{\VAN}[3]{##3}\VANthebibliography}

\usepackage{graphicx}	% Including figure files
\usepackage{amsmath}	% Advanced maths commands
\usepackage{amssymb}	% Extra maths symbols

\usepackage{newtxtext,newtxmath}

\newcommand{\noopsort}[1]{}
\renewcommand{\micron}{$\upmu$m}

\title[Collisions in a white dwarf planetary debris disc]{Collisions in a gas-rich white dwarf planetary debris disc}

\author[A. Swan et al.]
{Andrew Swan$^{1,2}$\thanks{E-mail: \href{mailto:a.swan.17@ucl.ac.uk}{a.swan.17@ucl.ac.uk}},
Scott J. Kenyon$^{3}$,
Jay Farihi$^{1,2}$,
Erik Dennihy$^{4}$,
Boris T. G{\"a}nsicke$^{5}$,
J.~J.~Hermes$^{6}$,
\newauthor
Carl Melis$^{7}$,
and Ted von Hippel$^{8}$
\\
$^{1}$Department of Physics~\& Astronomy, University College London, Gower Street, London~WC1E~6BT, UK\\
$^{2}$Visiting Astronomer at the Infrared Telescope Facility\\
$^{3}$Smithsonian Astrophysical Observatory, 60~Garden Street, Cambridge, MA~02138, USA\\
$^{4}$Gemini Observatory\,/\,NSF’s NOIRLab, Casilla~603, La Serena, Chile\\
$^{5}$Department of Physics, University of Warwick, Coventry~CV4~7AL, UK\\
$^{6}$Department of Astronomy \& Institute for Astrophysical Research, Boston University, 725~Commonwealth Ave., Boston, MA~02215, USA\\
$^{7}$Center for Astrophysics and Space Sciences, University of California, San Diego, CA~92093-0424, USA\\
$^{8}$Department of Physical Sciences, Embry-Riddle Aeronautical University, Daytona Beach, FL~32114, USA\\
}
\date{Accepted XXX. Received YYY; in original form ZZZ}

\pubyear{2021}

\begin{document}
\label{firstpage}
\pagerange{\pageref{firstpage}--\pageref{lastpage}}
\maketitle

\begin{abstract}
WD\,0145+234 is a white dwarf that is accreting metals from a circumstellar disc of planetary material. It has exhibited a substantial and sustained increase in {3--5\,\micron} flux since 2018. Follow-up \textit{Spitzer} photometry reveals that emission from the disc had begun to decrease by late 2019. Stochastic brightening events superimposed on the decline in brightness suggest the liberation of dust during collisional evolution of the circumstellar solids. A simple model is used to show that the observations are indeed consistent with ongoing collisions. Rare emission lines from circumstellar gas have been detected at this system, supporting the emerging picture of white dwarf debris discs as sites of collisional gas and dust production.
\end{abstract}


%%%%%%%%%%%%%%%%%%%%%%%%%%%%%%%%%%%%%%%%%%%%%%%%%%

%%%%%%%%%%%%%%%%% BODY OF PAPER %%%%%%%%%%%%%%%%%%

\section{Introduction}
\label{sectionIntroduction}

The host stars of virtually all known planetary systems will -- or have already -- become white dwarfs. Giant branch evolution and stellar mass loss may dramatically alter system architecture, but planetary material persists. Up to 50~per~cent of white dwarfs are polluted by metals accreted from orbiting rocky material, often with compositions strikingly similar to those of solar system objects \citep{Zuckerman2003, Zuckerman2010, Koester2014, Jura2014review}. Evidence for closely-orbiting planets is emerging \citep{Gansicke2019, Vanderburg2020planet}, and transits by dust clouds at several stars establish that circumstellar debris is commonplace \citep{Vanderburg2015, Vanderbosch2020, Guidry2021}.

Discs of hot dust are known from their excess infrared emission at around 1.5~per~cent of white dwarfs, and are suspected to exist undetected at many more \citep{Jura2007GD362, VonHippel2007DAZ, Farihi2008warmDust, Rocchetto2015, Farihi2016, Wilson2019}. In rare cases, at just 0.067~per~cent of white dwarfs, emission from gaseous debris accompanies dust \citep{Gansicke2006, Manser2020}, and recent evidence suggests collisions, as opposed to sublimation, are the underlying gas production mechanism \citep{Farihi2018GD56, Swan2020dust, Malamud2021}. Gas line profiles exhibit morphological variation consistent with precession of debris on eccentric orbits \citep{Manser2016doppler, Dennihy2018, Cauley2018, Miranda2018}, while short-period equivalent-width variations appear to be linked to closely-orbiting planetesimals \citep{Manser2019}. However, neither kind of variability is observed universally.

In 2018, an infrared outburst occurred at WD\,0145+234, a star where metals, dust, and gas have been detected \citep{Wang2019_0145, Melis2020gas}. The system brightened by over 1\,mag in the near-infrared within six months, the largest variation yet witnessed at any single white dwarf. However, infrared variability is widespread at white dwarf planetary systems, with a correlation between high variation amplitude and the presence of gaseous debris \citep{Xu2014drop, Farihi2018GD56, Xu2018IRvariability, Swan2019infrared, Swan2020dust}. WD\,0145+234 is thus at the high end of a spectrum of activity, but otherwise representative of its class.

The popular model for the origin of white dwarf metal pollution is that asteroids are perturbed onto eccentric orbits, tidally disrupted when they pass within the stellar Roche limit, and formed into a disc of closely-orbiting debris that is subsequently accreted \citep{Jura2003}. However, despite dozens of known examples, understanding of these discs is still evolving, and fundamental gaps remain. Numerous mechanisms have been proposed to perturb material onto nearly star-crossing orbits, of which only wide stellar companions have been shown to be unimportant \citep{Debes2002, Bonsor2011, Bonsor2015wideBinaries, Petrovich2017, Mustill2018, Wilson2019, Smallwood2021}. The details of subsequent disc formation are largely unconstrained, although theoretical efforts have produced plausible models \citep{Veras2014disruption, Veras2015shrinkingRings, Malamud2020A, Malamud2020B}, where interactions between solids and gas appear to play a key role \citep{Grishin2019, Rozner2021, Malamud2021}. Disc lifetimes are uncertain, with model predictions in the range $10^1$--$10^6$\,yr \citep{Rafikov2011PRdrag, Wyatt2014}, and observational estimates in the range $10^4$--$10^6$\,yr \citep{Girven2012, Cunningham2021}. Configurations, i.e. geometry and optical depth, are also uncertain. The model most often invoked envisages a compact, geometrically thin, optically thick disc \citep{Jura2003}, whose inner edge feeds accretion via Poynting--Robertson (PR) drag \citep{Rafikov2011PRdrag}. It is consistent with infrared fluxes at most systems, albeit with strong degeneracies in its parameters, but there are cases where it fails \citep{Jura2007Spitzer11stars, Farihi2017SDSS1557, GentileFusilloGasDiscs}. Moreover, it does not predict infrared variability, with the possible exception of disc depletion in runaway accretion events \citep{Rafikov2011runaway}. Optically thin models have also been proposed, and they can easily account for variation if dust is produced or destroyed \citep{Bonsor2017, Farihi2018GD56}. Each model on its own has disadvantages, but they are not mutually exclusive.

Collisions of solid particles within debris discs have been proposed to be responsible for infrared variability and evolution of the transit profiles of orbiting material \citep{Farihi2018Magnetism, Farihi2018GD56, Swan2020dust}, but there are other possible causes. If the eccentric debris stream produced by a tidal disruption is long-lived, Keplerian shear will redistribute material around the ring, imprinting periodic signals on the light curve as clumps are illuminated during perihelion passage \citep{Nixon2020}. Alternatively, dust liberated by collisions between a circularising debris stream and an existing disc may form a halo that evolves under PR~drag, potentially leading to flux variation on timescales of years \citep{Malamud2021}. At later stages of evolution, a compact, near-circular disc may grind material down into dust that sublimates and feeds accretion onto the star. That process is likely quiescent while in equilibrium, but stochastic delivery of fresh material can lead to variation in the gas production rate and infrared flux \citep{Kenyon2017collisions, Kenyon2017gas}. A combination of these mechanisms may generate observed light curves, but it is increasingly clear that collisions play an important role.

This paper reports post-outburst observations of WD\,0145+234 from the \textit{Spitzer Space Telescope} \citep{Werner2004Spitzer} and the NASA Infrared Telescope Facility (IRTF). Light curves, spectra, and derived quantities are presented in Section~\ref{sectionObservations}, shown to be consistent with a simple collisional model in Section~\ref{sectionCollisions}, and discussed in Section~\ref{sectionDiscussion}. A summary is given in Section~\ref{sectionSummary}.

\section{Observations and data analysis}
\label{sectionObservations}

The infrared outburst was announced less than four months before \textit{Spitzer} retired, so follow-up observations were urgently scheduled during the final visibility window for the target. Time-series imaging was acquired using both the 3.6- and {4.5-\micron} channels of the Infrared Array Camera (IRAC; \citealt{Fazio2004}), covering 15~epochs at a cadence of approximately 2\,d. During each visit, 12 frames per channel were acquired using 12-s exposures, dithered in the medium cycling pattern. The target was also observed once in an unrelated program, where 11 frames per channel were similarly dithered, with an exposure time of 30\,s. Observations from both programs span the period 2019~November~08 to 2019~December~27.

The \textit{Spitzer} data were reduced following the procedure detailed by \cite{Swan2020dust}, but a brief summary is given here. All frames were inspected for cosmic ray hits on the target star, as the automated masking in the mission pipeline is not guaranteed to be perfect; 8~out of 382 frames were discarded. Mosaics with 0\farcs6-square pixels were assembled from the calibrated images using \textsc{mopex}\footnote{\label{footnoteMopex}\href{https://irsa.ipac.caltech.edu/data/SPITZER/docs/dataanalysistools/tools/mopex/}{irsa.\hspace{0pt}ipac.\hspace{0pt}caltech.\hspace{0pt}edu/\hspace{0pt}data/\hspace{0pt}SPITZER/\hspace{0pt}docs/\hspace{0pt}dataanalysistools/\hspace{0pt}tools/\hspace{0pt}mopex/}}. Aperture photometry was then performed with \textsc{apex}\textsuperscript{\ref{footnoteMopex}}, using an aperture and sky annulus of 3 and 12--20 native pixels, respectively. The recommended array-location-dependent and aperture corrections were applied. Calibration uncertainties of 2~per~cent apply to absolute flux measurements \citep{Reach2005IRAC}, but were not added in quadrature to the photometric errors, as the data form a time series whose relative values are analysed.

While not pivotal to the analysis, photometry is used from \textit{WISE}, a space-based infrared observatory conducting an ongoing all-sky survey with a six-month cadence \citep{Wright2010}. Weighted mean fluxes and uncertainties were derived at each epoch, after discarding measurements flagged as problematic. The data used here are the same as relied upon for the initial discovery of the outburst \citep{Wang2019_0145}.

\begin{figure*}
\includegraphics[width=\textwidth]{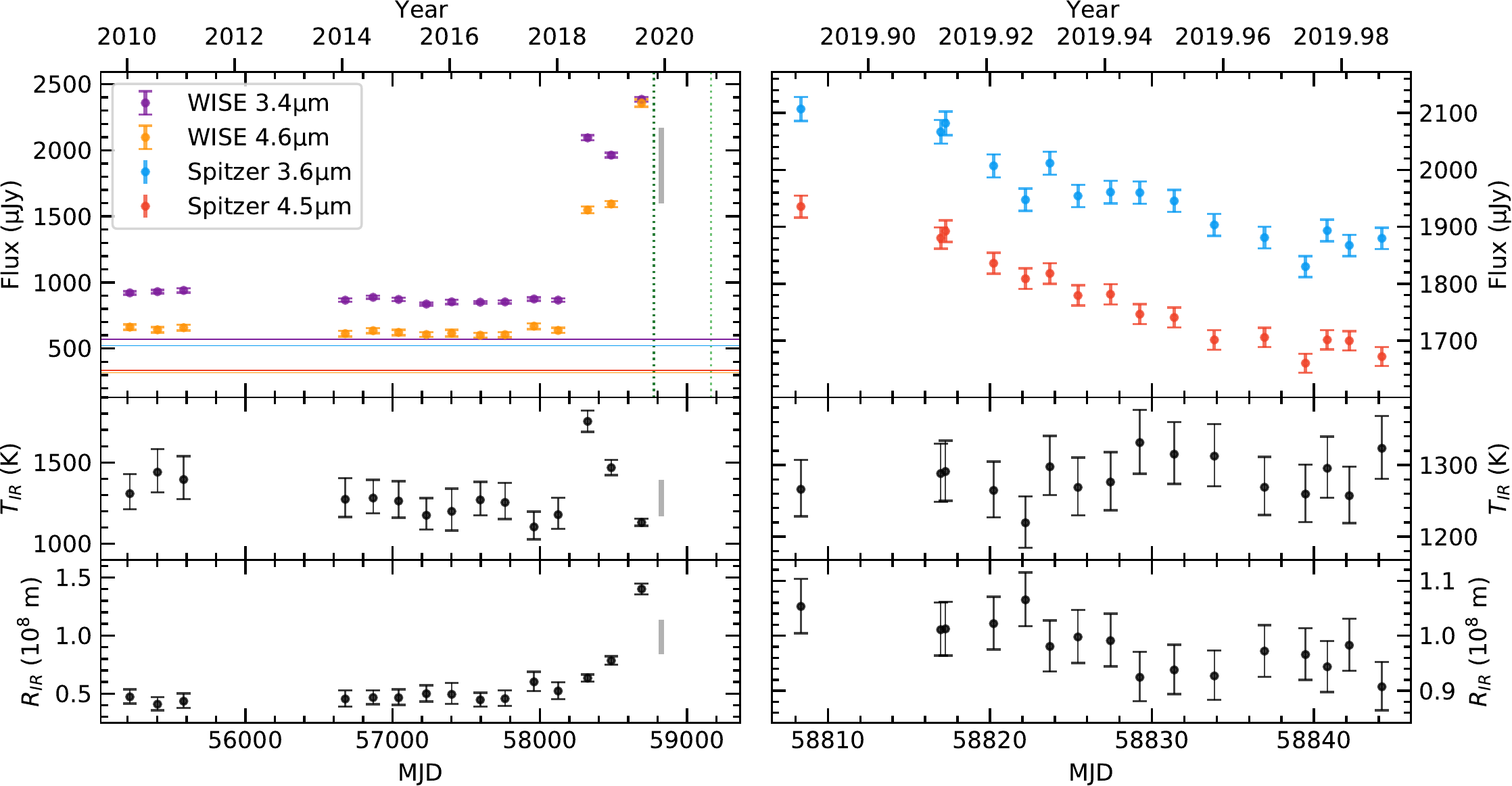}
\caption{Top panels: infrared light curve, where the right-hand panel shows a zoomed-in view. Vertical dotted lines indicate the acquisition dates of the SpeX spectra shown in Figure~\ref{figureIRTF}, and horizontal lines show the photospheric flux in each bandpass. Middle and bottom panels: temperature and radius of blackbody fit to the infrared excess at each epoch. Filled grey rectangles in the left-hand panels show the extents of the $x$- and $y$-axes of the right-hand panels.}
\label{figureLightcurve}
\end{figure*}

Figure~\ref{figureLightcurve} shows photometry from \textit{WISE} and \textit{Spitzer}, and clearly illustrates the outburst event, the onset of sustained colour changes that followed, and a subsequent decay in flux. In the panel displaying the \textit{Spitzer} observations, the downtrend is overlaid by stochastic brightening events correlated in both channels (e.g.~near MJD~58\,840), which are analysed in Section~\ref{sectionCollisions}.

Spectral energy distributions (SEDs) are constructed for each photometric epoch, and are used to model the excess infrared emission. A two-stage procedure is used, first inferring the stellar parameters ($T_{\text{eff}}$ and $\log{g}$) from optical photometry and a parallax measurement, then performing blackbody fits using the infrared data. This approach simplifies computation, and has a negligible impact on the blackbody parameter uncertainties. The priors, models, and data for each stage are now described, where uncertainties on all data are assumed to be Gaussian.

The stellar parameters ($T_{\text{eff}}$ and $\log{g}$) are inferred using Gaussian priors that are weighted averages of their reported values \citep{Gianninas2011, Limoges2015}, and a uniform prior on the distance. Photometry is retrieved from Pan-STARRS DR2 \citep{Chambers2019}, and assumed to be constant as the star has not been reported to be variable at optical wavelengths, despite its proximity to the ZZ\,Ceti instability strip. The stellar photosphere is modelled by interpolating over a grid of theoretical spectra for pure hydrogen atmospheres \footnote{Retrieved from the Spanish Virtual Observatory at \href{http://svo2.cab.inta-csic.es/theory/newov2/}{svo2.\hspace{0pt}cab.\hspace{0pt}inta-csic.es/\hspace{0pt}theory/\hspace{0pt}newov2/}} \citep{Koester2010}, extended with blackbody curves at long wavelengths, and the stellar radius is determined from the Montreal grids\footnote{\href{https://www.astro.umontreal.ca/~bergeron/CoolingModels/}{www.\hspace{0pt}astro.\hspace{0pt}umontreal.ca/\hspace{0pt}$\sim$bergeron/\hspace{0pt}CoolingModels/}} \citep{Bedard2020}. The distance is inferred from the \textit{Gaia}~EDR3 parallax \citep{Gaia2020}, and used to calculate the stellar fluxes in each bandpass.

To model the excess infrared emission, the stellar parameters are held constant at their median posterior values. The combined optical and infrared data are then fitted with the fixed stellar model plus a blackbody, whose parameters are inferred separately at each epoch. Priors are set on the blackbody temperature and radius using means and standard deviations for the dusty white dwarf population \citep{Rocchetto2015}. Values for the aforementioned quantities are summarised in Table~\ref{tableSEDfitting}, but note that the stellar parameters are subject to additional systematic uncertainties of at least 1.2~per~cent in $T_{\textrm{eff}}$ and 0.038\,dex in $\log{g}$ \citep{Liebert2005}. Results are shown in the lower panels of Figure~\ref{figureLightcurve}. There is considerable scatter in the median values, but the blackbody parameters at any given epoch are strongly correlated.

\begin{table}
%    \caption{Summary of SED fitting values and distributions. Posterior medians are given with formal uncertainties only.}
    \caption{Summary of distributions from SED fitting. Posterior medians are given with formal uncertainties only.}
    \centering
    \label{tableSEDfitting}
\begin{tabular}{ll}
\hline Quantity & Distribution/value \\
\hline
\textit{Priors}\\
\hspace{2mm}$P(\log{g\textrm{\,[cm\,s\textsuperscript{$-2$}]}})$ & $\sim\mathcal{N}(8.12,0.05)$ \\
\hspace{2mm}$P(T_{\text{eff}}\textrm{\,[K]})$                    & $\sim\mathcal{N}(12986,154)$ \\
\hspace{2mm}$P(\log{R_{\text{IR}}\textrm{\,[m]}})$               & $\sim\mathcal{N}(8.0,0.3)$   \\
\hspace{2mm}$P(T_{\text{IR}}\textrm{\,[K]})$                     & $\sim\mathcal{N}(1130,320)$  \\
%\textit{External data}\\
%\hspace{2mm}parallax\,(mas)                                      & $33.98\pm0.02$               \\
\textit{Posteriors}\\
\hspace{2mm}$T_{\textrm{eff}}$\,[K]                              & $12$\,$935\pm140$           \\
\hspace{2mm}$\log{g\textrm{\,[cm\,s\textsuperscript{$-2$}]}}$    & $8.07\pm0.01$               \\
\hline
\end{tabular}
\end{table}

\begin{figure}
\includegraphics[width=\columnwidth]{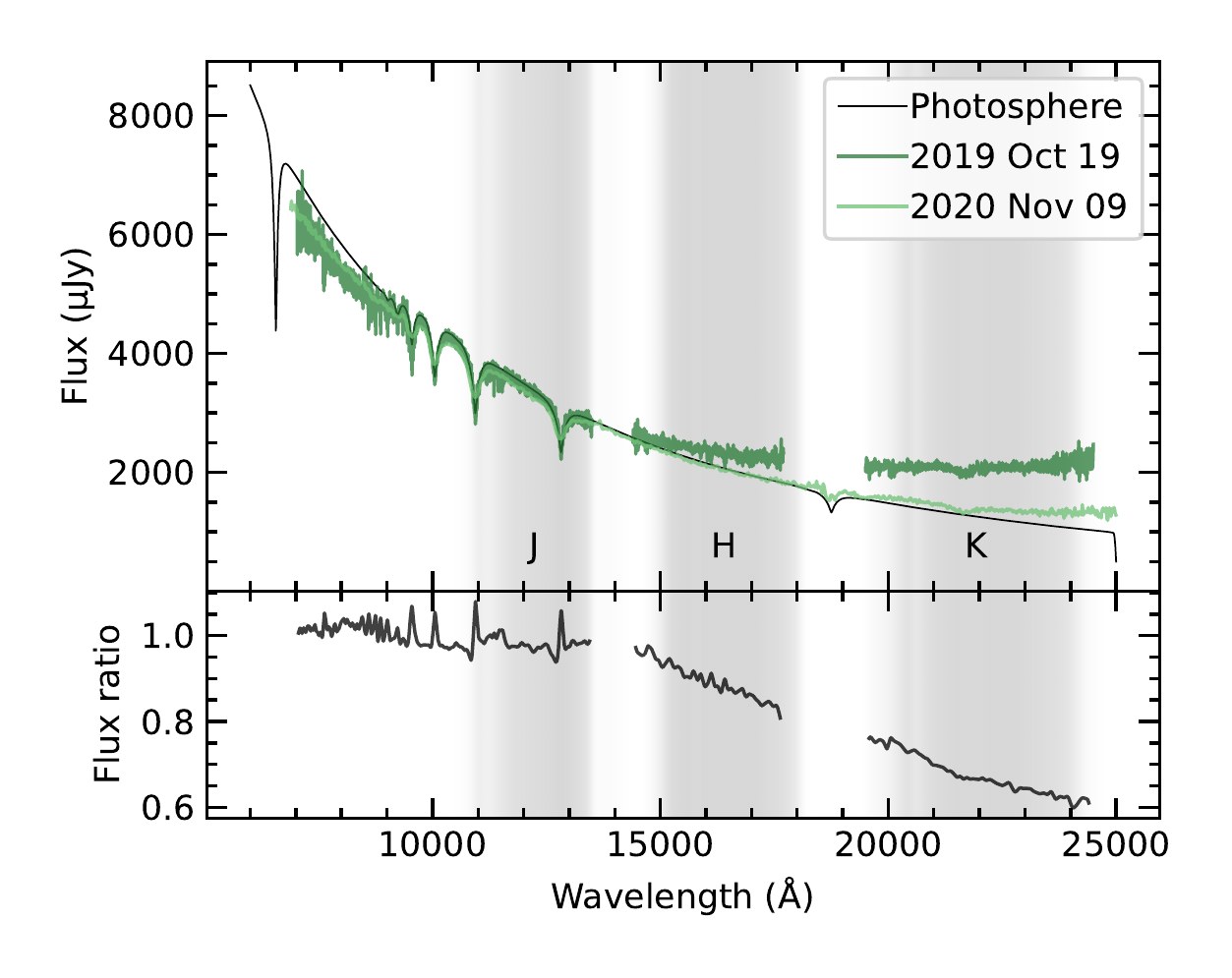}
\caption{Top panel: SpeX spectra taken approximately 1\,yr apart, overlaid on a model of the stellar photosphere interpolated from published grids \citep{Koester2010} using parameters found by fitting optical photometry (Table~\ref{tableSEDfitting}). Grey shading shows the \textit{JHK} passband profiles.
Bottom panel: ratio between 2020 and 2019 spectra highlighting the decline in \textit{H}- and \textit{K}-band flux.}
\label{figureIRTF}
\end{figure}

Spectroscopy was obtained on 2019~October~19 and 2020~November~09 using the SpeX instrument \citep{Rayner2003} on the IRTF at Mauna Kea, Hawaii. The 2019 spectra were taken in cross-dispersed mode with a 0\farcs5-slit, covering 0.70--{2.55\,\micron} at a resolving power of $R\simeq1200$, obtaining 70 exposures of 120\,s each. The 2020 spectra were taken using the prism and a 0\farcs8-slit, covering a similar wavelength range at $R\simeq200$, obtaining 38 exposures of 60\,s each. Standard procedures were followed: the telescope was nodded, arc lamp calibrations were employed, and a nearby A0V telluric standard star was observed. The data were reduced using \textsc{spextool}, producing wavelength- and flux-calibrated spectra \citep{Vacca2003, Cushing2004}. These are shown in Figure~\ref{figureIRTF}, together with a stellar photosphere model for the best-fit stellar parameters \citep{Koester2010}. SED fitting is attempted, but produces unreliable results, owing to imperfect flux calibrations and the lack of longer-wavelength data to constrain the blackbody temperature. However, it is clear from visual inspection that the spectra are consistent with a significant reduction in emitting area, in line with the trend in the photometry. The spectra cover the \ion{Ca}{ii} triplet near 8600\,\AA, a signature feature of circumstellar gas at white dwarfs. The triplet is detected in emission, but no analysis is attempted because of insufficient resolution, low signal-to-noise, and telluric standard correction artefacts. Higher-quality data that cover the triplet at a similar epoch have been published elsewhere \citep{Melis2020gas}.

\section{Collisional modelling of the light curve}
\label{sectionCollisions}

This section details a toy model for collisions within a debris disc, and its application to the decaying portion of the light curve observed by \textit{Spitzer}. It is emphasised that this is strictly a proof of concept, whose utility is confined to answering the question of whether the light curve is consistent with collisional evolution of circumstellar debris. It does not attempt a complete description of the system, nor does it address the initial rise in flux during the brightening phase of the outburst event.

Models of planetesimal belts undergoing collisional cascades predict that their total mass, and their infrared emission, will decrease with time. The form of the decay curve can be derived using simple arguments \citep{Dominik2003, Wyatt2007hotDust, Wyatt2007evolution}. In a disc containing $N$ objects, each with surface area $\sigma$, on orbits that fill a volume $V$ and intersect each other at a characteristic velocity $v$, the time for a single object to sweep out the whole volume is $t_{\text{s}}=\frac{V}{v\sigma}$. It will therefore encounter other objects on a timescale $t_{\text{coll}}=t_{\text{s}}/N$. Assuming such encounters are completely destructive, the population will diminish as:

\begin{equation}
\label{eqndNdtDerivation}
\frac{dN}{dt}=-2\frac{N^2}{t_s}
\end{equation}

This has a solution that sets a characteristic timescale $t_0$ for collisions among the population:

\begin{equation}
\label{eqnDecay}
N(t)=\frac{N_0}{1+t/t_0}\text{ where } t_0=\frac{t_s}{2N_0}
\end{equation}

For optically thin material, the emitted flux, mass, and surface area of the collisional swarm all follow equivalent relations. The flux remains approximately constant at early times ($t<t_0$), but on long timescales ($t \gg t_0$) it follows a $1/t$ curve.  Another consequence of Equation~\ref{eqnDecay} is that larger swarms process their material faster, i.e. all swarms follow the same curve at late times, so there is a degeneracy between $N_0$ and $t_0$ \citep{Wyatt2007evolution}.

The dimming in the \textit{Spitzer} light curve (Figure~\ref{figureLightcurve}) is consistent with such a decay. Overlaid on that trend are brightening events of amplitude $\sim1\%$ correlated in both channels, suggesting dust production in stochastic collisions. While the $1/t$ decay represents behaviour averaged over timescales $\delta t>t_0$, the stochastic component probes timescales comparable to $\delta t=t_0$. These brightening events are the key feature for which the following analysis seeks to account, where they are modelled as collisions between two objects. No attempt is made to model the initial rise in flux and subsequent decay that result from the collective behaviour of many solids, as a substantial numerical effort would be required to explore the wide parameter space. Instead, the aim of this initial study is to validate the collisional framework by focusing on this prominent and well-sampled aspect of it, and thereby determine whether more complex modelling is justified.

The post-outburst infrared excess may be emission from material liberated by a catastrophic collision between planetesimals. A simple model is now used to determine whether ongoing collisions within the resulting debris field can be responsible for the observed behaviour. By considering objects with a power-law size distribution that reside in a low-eccentricity annulus, the frequency of and emitting area liberated by collisions can be calculated and compared with observations. The frequency will be governed by the orbital parameters, the dimensions of the annulus, and the size of the objects. The emitting area will only depend on the size of the objects.

For simplicity, the excess infrared emission is assumed to come solely from optically thin material. The modest, stable pre-outburst infrared excess could be from debris that had settled into a flat disc configuration \citep{Jura2003}, but if it has survived, it would be only a minor component of the post-outburst flux.

Following \cite{Kenyon2016cascades}, the debris is assumed to occupy an elliptical annulus of semi-major axis $a$, radial width $\delta a$, and vertical extent $H$, so that its volume is $V=4\pi a\delta aH$. Orbits within the annulus are characterised by Keplerian velocity $v_{\text{K}}=\sqrt{GM_{\star}/a}$ and period $P=2\pi/\Omega_{\text{K}}$, where $G$ is the gravitational constant, $M_{\star}$ is the stellar mass, and $\Omega_{\text{K}}$ is angular frequency. Where particles have typical eccentricity $e$ and inclination $i\approx e/2$, they will encounter each other with relative velocity $v_{\text{rel}}\approx ev_{\text{K}}$, and have vertical velocity $v_{\text{z}}\approx iv_{\text{K}}$. The vertical extent of the annulus can be approximated by $H=v_{\text{z}}/\Omega_{\text{K}}$. Its semi-major axis is estimated from the optical depth $\uptau$, taken to be the covering fraction, i.e. the fraction of starlight intercepted and reprocessed by dust with total surface area $A$:

\begin{equation}
\label{eqnTau}
a=\sqrt{A/4\pi \tau}
\end{equation}

Objects are assumed to have a density $\rho=3$\,g\,cm\textsuperscript{$-3$} (representative of solar system asteroids; \citealt{Carry2012}), and to form a collisional swarm, with a power-law size distribution $n(r)\propto r^{-q}$ where $r$ is the object radius and the exponent $q\approx3.5$ \citep{Dohnanyi1969}. Integrating over the size distribution provides expressions for the total number of objects $N$, their collective surface area $A$ (as defined in the previous paragraph), and their total mass $M$ between object sizes $r_1$ and $r_2$:

\begin{equation}
\label{eqnN}
N=N_0\int_{r_1}^{r_2} r^{-q} \,dr
\end{equation}

\begin{equation}
\label{eqnA}
A=N_0\int_{r_1}^{r_2} \pi r^{2-q} \,dr
\end{equation}

\begin{equation}
\label{eqnM}
M=N_0\int_{r_1}^{r_2} \textstyle\frac{4}{3}\pi\rho r^{3-q} \,dr
\end{equation}

The constant of proportionality $N_0$ can be determined given one of those quantities, e.g.:

\begin{equation}
\label{eqnN0}
N_0 = \frac{(3-q)A}{\pi(r_2^{3-q}-r_1^{3-q})} = \frac{(4-q)M}{\textstyle\frac{4}{3}\pi\rho(r_2^{4-q}-r_1^{4-q})}
\end{equation}

For power law slopes in the range of interest ($3<q<4$), one of the $r_1$ and $r_2$ terms in the denominator becomes negligible in each form of Equation~\ref{eqnN0}, but dominates in the other form. This apparent contradiction is a reflection of the fact that most of the mass is contained in large objects, and thus governed by $r_2$, while the smallest grains account for most of the emitting area, governed by $r_1$. The observed infrared emission only constrains $A$, and thus the size of the largest object in the swarm (and total swarm mass) can be set to an arbitrarily large value without significantly affecting the results.

It is important to appreciate that the collisional cascade continually creates and removes emitting surface area. Objects of any given size are destroyed in collisions, but replenished by collisions between larger objects, in an equilibrium that maintains the power law distribution. The smallest dust grains are vapourised when they collide (i.e. $r_1$ remains constant). The long-term decay arises because the largest objects cannot be replenished (i.e. $r_2$ decreases when the largest object in the swarm suffers a collision). It is equally important to appreciate that this process is stochastic: large objects collide infrequently, producing a lot of dust, while smaller objects will collide more often, but produce less dust. The contributions of collisions more frequent than the observation cadence are smoothed out in a light curve, but the larger, less frequent collisions are not, and thus leave a detectable signature.

Now that the model has been defined, the question is whether it can account for the observed light curve. The median time between epochs is 2.0\,d, during which time flux increases of $\sim1\%$ sometimes occur. If these events represent stochastic, destructive encounters between objects in the swarm, then there should be a characteristic object size that has an expected rate of collisions of around 0.5\,$\text{d}^{-1}$, where each collision liberates sufficient material to increase the (instantaneous) total surface area by $\sim1\%$. If so, that will permit an order-of-magnitude estimate of the parameters of the observed collisions.

Dust production is most efficient in destructive encounters between objects of similar sizes. The steep power law means that a given object will encounter objects of a similar size to itself much more often than larger objects, while it is unlikely to be destroyed by more frequent impacts from particles much smaller than itself. To simplify the analysis, therefore, all collisions are treated as fully destructive, but those involving size ratios significantly different to unity are ignored. In practical terms this is achieved by dividing the size distribution into bins, and considering only those collisions that occur between objects in the same bin.

The population of a given size bin will be depleted by collisions, and replenished by fragments produced in larger bins, so that $\dot{N}=\dot{N}_{\text{repl}}+\dot{N}_{\text{coll}}$ within that bin. In the largest bin, $\dot{N}_{\text{repl}}=0$, but for all others, $\dot{N}_{\text{repl}}\sim-\dot{N}_{\text{coll}}$. The rate of collisions can be calculated by considering the density of objects, their cross-sections, and their relative velocities. A single object of size $r$ will experience collisions with the $N$ objects of size $R$ within its bin at a rate:

\begin{equation}
\label{eqnCollisionRateSingleObject}
\mathcal{R}_{\text{coll}} =\frac{N}{V}\pi(r^2+R^2)v_{\text{rel}}
\end{equation}

This is similar to calculating the sweep time $t_s$ mentioned at the beginning of this section. However, Equation~\ref{eqndNdtDerivation} assumes $N$ is large, which is not valid for the larger size bins considered here. Naively, $\dot{N}_{\text{coll}}=-\mathcal{R}N\propto N^2$, but objects cannot collide with themselves, and double-counting collisions between each pair must be avoided, so a factor $N(N-1)/2$ replaces $N^2$. The number of collisions in time $dt$ amongst all $N$ objects within a single bin where $r\approx R$ is thus:

\begin{equation}
\label{eqndN}
dN_{\text{coll}} =\frac{N(N-1)}{2V} \pi R^2 v_{\text{rel}}\,dt
\end{equation}

Note that the collision rate is independent of eccentricity: from the model specification above, $V \propto H \propto v_{\text{z}} \propto i \propto e$ and $v_{\text{rel}} \propto e$, so the the terms in Equation \ref{eqnCollisionRateSingleObject} cancel. In reality, the radial extent of the annulus $\delta a$ (on which $V$ depends) is determined by eccentricity as well as the spread in semi-major axes. However, this is not a concern in this toy model, as even small eccentricities of $e\sim0.01$ lead to relative velocities of several km\,s\textsuperscript{$-1$}, so that collisions can be assumed to be completely destructive.

Each collision is assumed to produce fragments that are no larger than $f$ times the mass of the parent objects, and that respect the power law. Taking $f=0.01$, the additional emitting area liberated by the collision can then be determined from Equations~\ref{eqnA} and~\ref{eqnN0}. %NOTE: this is reasonable because the fragments will be rapidly processed by and absorbed into the swarm.

To apply the model to WD\,0145+234, the results of the SED fitting described in Section~\ref{sectionObservations} are used to set values for system parameters. The stellar mass $M_{\star}=0.65$\,$\text{M}_{\odot}$ is interpolated from the Montreal grids \citep{Tremblay2011}. The blackbody area and effective temperature from the first epoch of \textit{Spitzer} photometry are used to evaluate Equations~\ref{eqnN} and~\ref{eqndN} over a suitable range of object sizes. Because small dust grains radiate inefficiently at wavelengths longer than their size, a lower bound is set at $r_{\text{min}}=0.3$\,\micron. An arbitrary upper bound is set at $r_{\text{max}}=100$\,km, chosen to be significantly larger than the total quantity of emitting material. Given the optical depth (covering fraction) from SED fitting of $\tau=0.016$, Equation~\ref{eqnTau} yields an estimate for the orbital semi-major axis of $a=9\times10^8$\,m. A small eccentricity $e=0.01$ is adopted, yielding a vertical extent $H=4\times10^6$\,m for the debris annulus, but it is reiterated that the model is largely independent of eccentricity.

\begin{figure}
\includegraphics[width=\columnwidth]{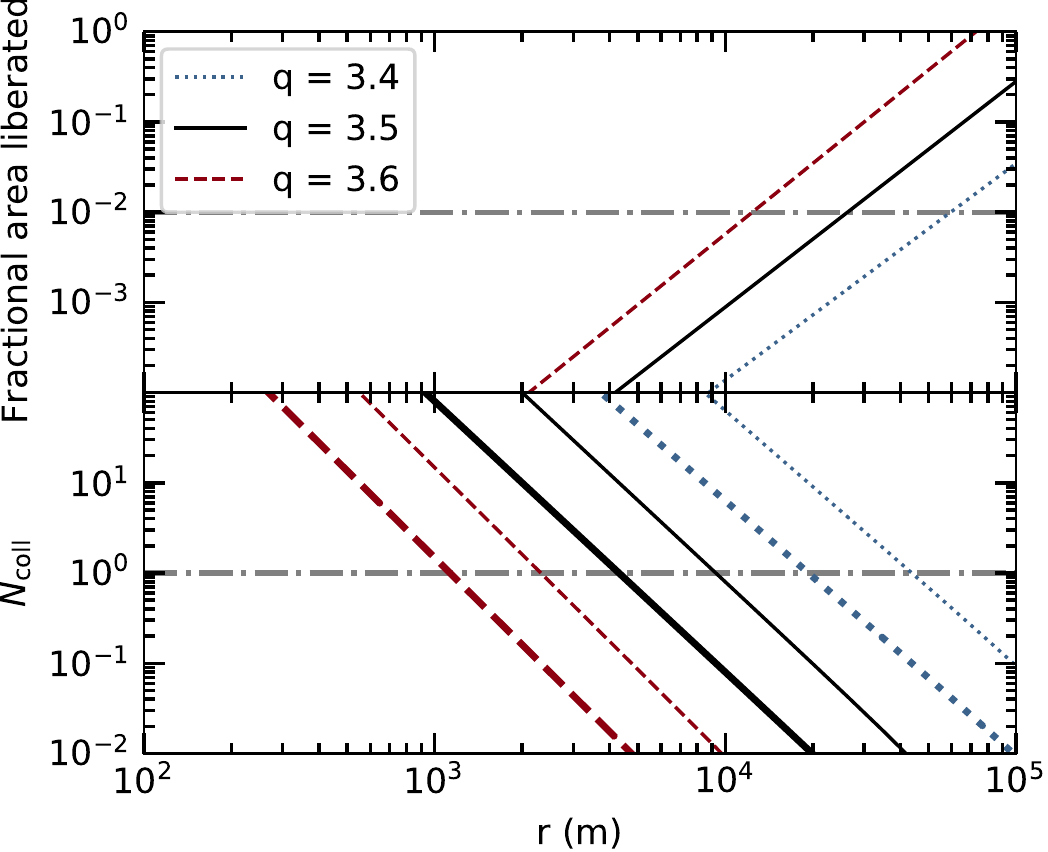}
\caption{Top panel: fractional emitting area of debris that is liberated by a single collision between objects of size $r$, shown for three different slopes ($q$) of the size distribution power law (there is no dependence on $\delta a/a$). Bottom panel: number of such collisions ($N_{\text{coll}}$) per time $dt=2$\,d, shown for three different power law slopes and two different fractional annulus widths (thick lines for $\delta a/a=0.01$, thin lines for $\delta a/a=0.001$). In each panel the dot-dashed grey lines represent the brightening events seen in the light curve. For example, if $q=3.5$ and $\delta a/a=0.01$, one collision every 2\,d can be expected between objects with radii of around 4\,km (lower panel), increasing the emitting area by 0.01\% (upper panel).}
\label{figureCollisionParameters}
\end{figure}

Collision rates are calculated, as are emitting areas liberated per collision, setting $dt=2$\,d, for 100 logarithmically-spaced size bins between $r_{\text{min}}$ and $r_{\text{max}}$. Results for different fractional annulus widths ($\delta a/a$) and power law slopes ($q$) are shown in Figure~\ref{figureCollisionParameters}. The horizontal dot-dashed lines indicate values that relate to the observations, i.e. one collision per 2\,d, generating sufficient additional surface area to increase the infrared flux by about 1\%. As the width of the annulus decreases, collisions between large objects become more frequent. A steeper power law has the opposite effect, as more mass is concentrated into smaller particles. While the value of the fragmentation parameter $f$ is chosen arbitrarily, it has minimal impact on the results: $f$ must be decreased by six orders of magnitude to cause the same change in dust area liberated as an increase in $q$ of 0.1.

\begin{figure}
\includegraphics[width=\columnwidth]{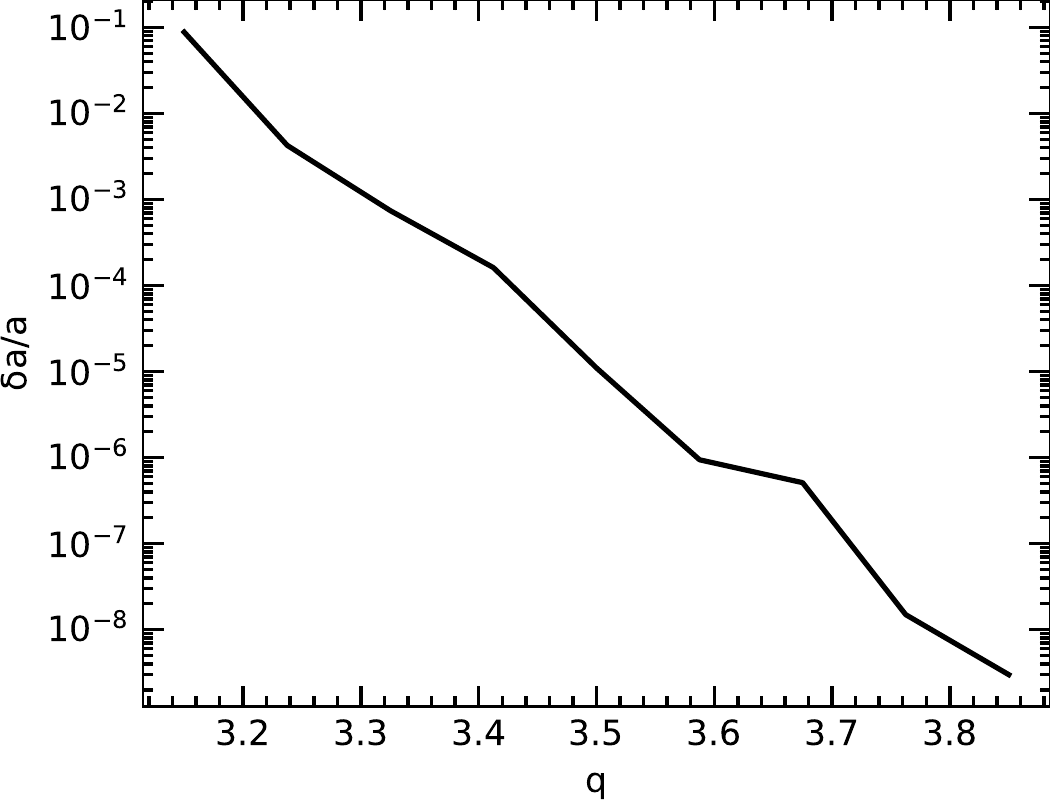}
\caption{Combinations of size distribution power law slope ($q$) and fractional annulus width ($\delta a/a$) that result in destructive encounters once every 2\,d, and liberate sufficient material to increase the observed flux by 1\%. The line is not smooth because of the size bins used in the calculations.}
\label{figureCollisionsQvsDeltaA}
\end{figure}

It is clear that reasonable choices for the model parameters can be made that produce collisions at the expected rate and scale. However, to be consistent with the observations, the lines in both panels of Figure~\ref{figureCollisionParameters} should cross the dot-dashed lines at the same $r$. In other words, there should be a characteristic size where objects experience collisions that liberate the observed quantity of dust on timescales approximately equal to the observed cadence. Parameter values that achieve this are now calculated. Equation~\ref{eqnN0} can be rearranged to find the largest fragment in the debris from a collision, for a given value of $q$, and thus the size of the parent body. That identifies the size bin for which collisions of interest occur, and Equation~\ref{eqndN} can then be used to find the value of $\delta a$ that yields $dN_{\text{coll}}=1$. Figure~\ref{figureCollisionsQvsDeltaA} shows the parameter combinations that produce collisions consistent with the observations.

The parameters of the collisional swarm can be compatible with the data across a wide range of values, but some choices are physically unlikely. Lower values of $q$ require higher values of $r_{\text{max}}$ in order to provide sufficiently large colliding objects, although the value of that parameter does not otherwise impact the results significantly. For example, setting $r_{\text{max}}=10^4$\,m requires that $q\gtrsim3.60$, while $r_{\text{max}}=10^5$\,m allows $q\gtrsim3.35$. Higher values of $q$ lead to unrealistically narrow annuli of debris: Figure~\ref{figureCollisionsQvsDeltaA} shows that for $q=3.7$, objects have orbits that differ by only $\delta a\approx 10^2$\,m (recalling that $a\approx 10^9$\,m). However, a value near the canonical $q=3.5$ yields a more reasonable $\delta a\approx 10^4$\,m. In that case, collisions between objects of $r\approx 25$\,km can liberate material with 1\,per~cent of the total emitting area every 2\,d, and it therefore appears that the model can account for the observations.

Stepping away from the model, a crude estimate can be made of the scale of the event that triggered the outburst. Equation~\ref{eqnA}, with $q=3.5$ and fragmentation parameter $f=0.01$, suggests that a collision between two objects of $r\approx 200$\,km would generate an emitting area sufficient to produce the peak observed flux. Such objects would have a combined mass of $M\approx 2\times 10^{20}$\,g. This is larger than previous estimates of the minimum dust mass \citep{Wang2019_0145}, because, as always, the size distribution concentrates mass in the largest objects, which contribute little emitting area. It is also undoubtedly an overestimate, as the peak flux occurred at least 1\,yr after the initial outburst, during which time continuing collisions within the debris would have generated more dust. Consideration of the duty cycle and rates of white dwarf metal accretion, minimum photospheric metal masses, and estimated disc lifetimes implies that objects large enough to cause the outburst event are likely to be readily available to white dwarf debris discs (e.g. \citealt{Koester2014, Hollands2018analysis}). Therefore, given the success of the simple analysis described here, a more detailed exploration of the outburst event may be worthwhile, extending the model to account for the initial rise in infrared emission.

\section{Discussion and outlook}
\label{sectionDiscussion}

The analysis in Section~\ref{sectionCollisions} establishes that the \textit{Spitzer} data presented in Figure~\ref{figureLightcurve} are consistent with ongoing collisions within the circumstellar material at WD\,0145+234. The long-term light curve is now interpreted in that context, and compared to that at another system. The nature of the infrared outburst is then examined, and the link with gas emission lines is discussed. Finally, the limitations of the model and some directions for future work are set out.

A modest, stable infrared excess apparently existed when observations began in 2010, persisting until the outburst event in 2018. That emission may have come from a compact, optically thick disc. The subsequent increase in both temperature and emitting area is consistent with the production of optically thin dust in a destructive collision, as previously noted \citep{Wang2019_0145}. Unshielded dust has a higher temperature at a given distance from the star than it would in an optically thick configuration, so a change in radial location is not necessarily implied; in other words, the collision could have happened within an existing debris disc, as envisaged in some models \citep{Kenyon2017collisions}. The emitting area increased towards a peak in 2019, while the temperature returned to pre-outburst values. If the colliding material is radially distinct from the original disc, the light curve could show it settling into an optically thick configuration, accompanied by radial spreading. That would account for the temperature decrease as dust becomes shielded from starlight, while angular momentum exchange during collisions increases the disc surface area, and thus the emitted flux. Alternatively, if the colliding material resides on orbits that intersect the original disc, debris liberated in ongoing collisions is likely to be absorbed into the disc on orbital timescales \citep{Farihi2018GD56}. In that scenario, the effective temperature traces the ratio of optically thin to optically thick emission, and therefore the net rate of dust production and destruction. However, at least some of the debris is likely to be scattered onto orbits beyond the disc, complicating this picture \citep{Malamud2021}.

\begin{figure}
\includegraphics[width=\columnwidth]{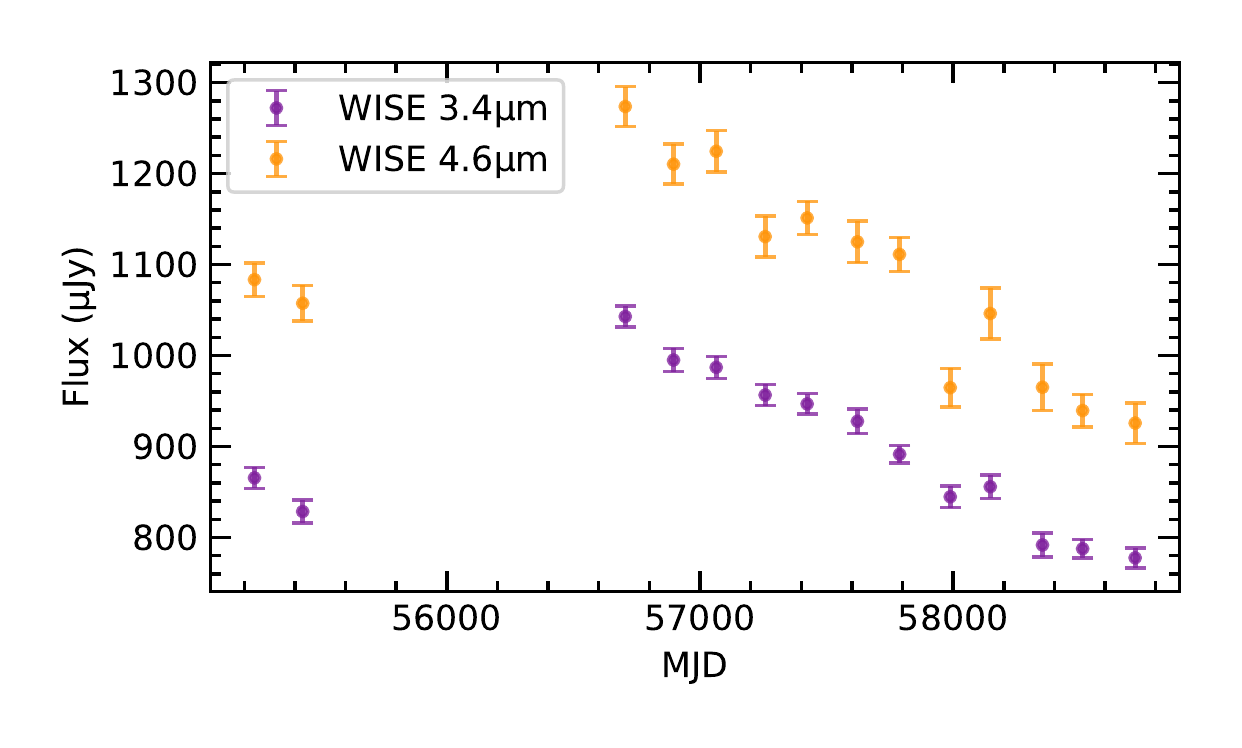}
\caption{Light curve for GD\,56, which resembles the \textit{Spitzer} data presented in Figure~\ref{figureLightcurve}, but on a timescale two orders of magnitude longer. The stellar photosphere contributes 120\,$\upmu$Jy and 65\,$\upmu$Jy in the \textit{WISE} \textit{W1}~(3.4\,\micron) and \textit{W2}~(4.6\,\micron) bands, respectively.}
\label{figureLightcurveGD56}
\end{figure}

The downtrend in the \textit{Spitzer} light curve could result from a change in surface area or temperature of the emitting material. Linear fits to the blackbody parameters inferred in Section\,\ref{sectionObservations} are consistent with a decrease in emitting area ($2.8\upsigma$ significance), but no change in temperature. Interestingly, the flux decay at WD\,0145+234 bears a striking resemblance to that identified at GD\,56 \citep{Farihi2018GD56}, where the latest \textit{WISE} data for that star show the trend continuing (Figure\,\ref{figureLightcurveGD56}). Despite timescales that differ by two orders of magnitude, both light curves are consistent with $1/t$ decay. Moreover, they are both overlaid with brightening events (e.g. near MJD~58\,000 for GD\,56). As argued above, all of these features are consistent with dust production and destruction in a collisional cascade, lending strong support to the idea that collisions are a primary cause of white dwarf debris disc evolution.

The infrared outburst has been interpreted as the tidal disruption of an asteroid \citep{Wang2019_0145}. However, such events occur at most once in the history of an object, requiring observation at a special time, whereas collisions within debris already orbiting close to the star are an ongoing process. WD\,0145+234 thus stands apart from other dusty white dwarfs only in the amplitude of its infrared variation, suggesting that it is merely the most active member of its class at present. While the infrared outburst motivated the high-cadence \textit{Spitzer} observations that enable this study, such an event is not a condition of the collisional model, which deals only with stochastic collisions within a population of planetesimals. Such a model is thus universally applicable to white dwarf debris discs.

Gaseous debris is observed at a subset of dusty white dwarfs, including WD\,0145+234. The results presented here reinforce the correlation between infrared variation and emission from gas, where collisions are proposed both as the cause of variation and as a source of gas \citep{Kenyon2017gas, Swan2020dust, Malamud2021}. Recondensation of gas onto dust grains provides a sink, and while there is not yet a theoretical consensus, that process potentially operates on orbital timescales \citep{Johnson2012, Metzger2012}. %Johnson eqn 13 is the relevant one, gives answer of only a few seconds at 100% sticking efficiency
Equivalent widths of emission lines in observed systems have typically remained stable over timescales longer than this, suggesting a steady state between gas production and recondensation. There are exceptions, however, where significant increases and decreases in emission line equivalent widths have been seen over months or more \citep{Wilson2014, Dennihy2020gas, GentileFusilloGasDiscs}. Under the model considered here, equilibrium may be maintained by constant gas production in frequent collisions between small objects, and disturbed by stochastic collisions between large objects. Should the rate of collisional gas production increase on a timescale shorter than that of recondensation, there is the potential for gas to accumulate temporarily. That may lead to an increase in line equivalent widths, unless the gas is optically thick, in which case the process will be hidden from view, and recondensation may be enhanced in the lower-temperature environment. Monitoring gas-emission systems for equivalent-width variation, or otherwise, with simultaneous infrared coverage, will be useful to inform modelling of gas production and recondensation.

In addition to long-term morphological variations consistent with precession, emission lines in the prototype gas emission system SDSS\,J122859.93+104032.9 also exhibit sinusoidal variations in equivalent width on a timescale of hours. These are proposed to trace the orbital period of a planetesimal embedded in a disc, where collisions generate the observed gas \citep{Manser2019, Manser2020}. The planetesimal size is not tightly constrained, but could be as small as 4\,km. That model is conceptually similar to that presented here~-- debris discs are by definition planetesimals embedded in a disc --~but it focuses on ongoing interactions with a single object, rather than numerous destructive collisions between similar-sized planetesimals among a population, as considered here. However, it is not suggested that there is any tension, as the model here is generic, and subsumes the special case where a single object becomes dominant owing to, say, possessing significantly higher internal strength than the surrounding debris.

The collisional model adequately accounts for the observed infrared variability, but some aspects of white dwarf debris discs remain unexplained. While the strongest infrared variation is accompanied by gas emission lines, as exemplified by WD\,0145+234, the correlation is not perfect \citep{Swan2020dust}. GD\,56 has a strikingly similar infrared light curve to WD\,0145+234, as discussed above, yet despite that strong variation it has never been reported to show emission from gas, including during the ongoing infrared dimming \citep{Farihi2018GD56}. Conversely, some systems with emission features have not displayed strong infrared variation, with the caveat that data are sparse. Intriguingly, post-outburst observations of the gas emission at WD\,0145+234 reveal no equivalent-width variations \citep{Melis2020gas}, and a similar lack of variation accompanied the infrared dimming event at WD\,J0959$-$0200 \citep{Xu2014drop}. The apparent absence of a correlation between infrared and equivalent-width variation may be an optical depth effect: if the emitting gas is optically thick, its production or destruction will cause no variation in equivalent width. Yet other systems display significant equivalent-width variability, most strikingly at SDSS\,J161717.04+162022.4, where the emission features first strengthened, then faded completely over several years \citep{Wilson2014}. That event was interpreted as an impact onto an existing disc, an idea that had previously been explored theoretically \citep{Jura2008manySmallAsteroids}. The persistence of gas beyond the likely short timescale for recondensation suggests a stream of debris would be responsible, rather than a single object, in line with recent work \citep{Malamud2021}. Such a configuration is distinct from the single annulus considered here, but the collisional model could be extended to explore that scenario, as the same principles apply.

The data and analysis presented here each have some limitations. The toy model assumes that all dust produced by collisions is optically thin, which may not be true initially, though spreading and Keplerian shear will quickly disperse it. Destructive collisions are assumed, but the model is independent of eccentricity, so collision velocities are unconstrained. However, that is not a concern: an eccentricity of $e\gtrsim0.002$ is sufficient for the model to work, found by equating the specific centre-of-mass collision energy of two equal-mass objects with the specific energy required for catastrophic destruction (\citealt{Kenyon2017collisions}, section~3.2).

The data cover only a small part of the relevant wavelength range, hence the choice of a single-temperature blackbody model in SED fitting. Wider spectral coverage would allow more detailed disc models to be tested, where longer wavelengths constrain the eccentricity of orbiting debris, and higher resolution allows an opaque disc to be distinguished from optically thin dust. Increased time resolution, as well as longer baselines, would reveal whether the stochastic brightening events seen here on a 2-d cadence also occur on other timescales, allowing a more stringent test of the collisional model. The data are at present insufficient to search for periodic changes that could be linked to orbital timescales.

The clear change between the two SpeX spectra makes WD\,0145+234 the third non-pulsating white dwarf known to vary in the \textit{K} band \citep{Xu2014drop, Xu2018IRvariability}. A three-year monitoring campaign found no significant \textit{K}-band variations among a sample of 34~stars, but would undoubtedly have been sensitive to the flux change reported here \citep{Rogers2020}. Ground-based monitoring of the dusty population will therefore continue to be valuable, as it has the potential to detect major collisional events at dusty stars. With \textit{Spitzer} no longer operational, the ongoing NEOWISE survey is the only space-based facility that will monitor this star in the infrared. However, \textit{JWST} and \textit{SPHEREx} should become operational in the next few years, promising enhancements in wavelength coverage and spectral resolution that will enable the investigations outlined above.

\section{Summary}
\label{sectionSummary}

This paper presents \textit{Spitzer} observations of WD\,0145+234 in the aftermath of its infrared outburst event. The light curve shows a decay trend suggestive of collisional evolution, and a toy model is employed to test whether the \textit{Spitzer} data are consistent with that scenario. An annulus of debris undergoing a collisional cascade is considered, whose constituent planetesimals follow a power-law size distribution. Encounters between similar-sized objects produce fresh debris, where a given object size results in collisions with a characteristic frequency and dust quantity. The model parameters are tuned to generate events that are consistent with the short-term variability in the infrared light curve, i.e. the small bumps superimposed on the decay trend. Reasonable choices of parameters can be made that fit the data, so the model is successful.

This analysis shows that the post-outburst light curve is consistent with collisions between planetesimals in a debris disc. Furthermore, the stable pre-outburst infrared excess suggests that a debris disc was already in place, whose mass of emitting dust grains increased during the event. Though neither the brightening phase of the outburst nor the origins of the proposed annulus of debris are modelled, it appears that interactions within circumstellar debris can account for the observations, without the need to invoke a tidal disruption. That conclusion is underlined by the similarity between the light curves at WD\,0145+234 and GD\,56, showing that such behaviour is not rare.

Emission from circumstellar gas has been detected at WD\,0145+234, consistent with the idea that planetesimal collisions drive both dust and gas production. However, it is not yet clear how the variation observed in infrared flux and in equivalent widths of emission lines are linked. While the source of gas appears to be collisions, sinks such as recondensation are less well understood, suggesting a direction for future work.

\section*{Acknowledgements}
%The Acknowledgements section is not numbered. Here you can thank helpful colleagues, acknowledge funding agencies, telescopes and facilities used etc. Try to keep it short.
The authors thank the referee Evgeni Grishin for a thoughtful and thorough review, and Uri Malamud for feedback on a draft. AS and JF thank Mike Connelly for local and remote observing assistance. AS~acknowledges support from a Science and Technology Facilities Council (STFC) studentship. JF~acknowledges support from STFC grant ST/R000476/1. ED~acknowledges support by NOIRLab, which is managed by the Association of Universities for Research in Astronomy (AURA) under a cooperative agreement with the National Science Foundation (NSF). BTG~was supported by STFC grant ST/T000406/1 and by a Leverhulme Research Fellowship. CM~acknowledges support from NSF grant SPG-1826583. TvH~acknowledges NSF grant AST-1715718. This research made use of \href{https://github.com/dfm/emcee}{emcee} \citep{Foreman-Mackey2013} and \href{https://www.astropy.org}{Astropy} \citep{Astropy2013, Astropy2018}. This work is based on observations made with the \textit{Spitzer Space Telescope}, which was operated by the Jet Propulsion Laboratory (JPL), California Institute of Technology (CIT) under a contract with NASA, and the Infrared Telescope Facility, which is operated by the University of Hawaii under contract 80HQTR19D0030 with NASA. This work has made use of data from the European Space Agency mission \href{https://www.cosmos.esa.int/gaia}{\it Gaia}, and the \href{https://panstarrs.stsci.edu/}{Pan-STARRS1 Surveys}. This publication makes use of data products from the \textit{Wide-field Infrared Survey Explorer}, a joint project of the University of California, Los Angeles and JPL/CIT, and NEOWISE, a project of JPL/CIT, both funded by NASA.

%%%%%%%%%%%%%%%%%%%%%%%%%%%%%%%%%%%%%%%%%%%%%%%%%%
\section*{Data Availability}
%The inclusion of a Data Availability Statement is a requirement for articles published in MNRAS. Data Availability Statements provide a standardised format for readers to understand the availability of data underlying the research results described in the article. The statement may refer to original data generated in the course of the study or to third-party data analysed in the article. The statement should describe and provide means of access, where possible, by linking to the data or providing the required accession numbers for the relevant databases or DOIs.
\textit{Spitzer} observations from programs 14220 (PI: Siyi Xu) and 14322 (PI: Andrew Swan) are stored in the \href{https://sha.ipac.caltech.edu/applications/Spitzer/SHA}{Spitzer Heritage Archive}. IRTF observations from programs 2019B030 and 2020B050 (PI: Jay Farihi) will be available from the NASA/IPAC \href{https://irsa.ipac.caltech.edu}{Infrared Science Archive} (IRSA) after an 18-month proprietary period. \textit{WISE} data are also available from \href{https://irsa.ipac.caltech.edu}{IRSA}.



\bibliographystyle{mnras}
\bibliography{WD0145+234}

%\begin{thebibliography}{99}
%\end{thebibliography}

%%%%%%%%%%%%%%%%%%%%%%%%%%%%%%%%%%%%%%%%%%%%%%%%%%

%%%%%%%%%%%%%%%%% APPENDICES %%%%%%%%%%%%%%%%%%%%%

%\appendix


%If you want to present additional material which would interrupt the flow of the main paper, it can be placed in an Appendix which appears after the list of references.

%%%%%%%%%%%%%%%%%%%%%%%%%%%%%%%%%%%%%%%%%%%%%%%%%%

% Don't change these lines
\bsp	% typesetting comment
\label{lastpage}
\end{document}